\documentclass[aps,pre,reprint,10pt,twocolumn,showpacs,nofootinbib,superscriptaddress]{revtex4-1}

\usepackage{amsmath}
\usepackage{amsfonts}
\usepackage{amssymb}
\usepackage[pdftex]{graphicx}
\usepackage[pdfborder={0 0 0}]{hyperref}  
\usepackage{color}
\usepackage[latin1]{inputenc}
\usepackage{float}
\usepackage{comment}

\newcommand{\lrangle}[1]{\langle {#1} \rangle}

\begin{document}
\title{Sampling methods for the quasistationary regime of epidemic processes on regular and complex networks}
 
\begin{abstract}
A major hurdle in the simulation of the steady state of epidemic processes is
that the system will unavoidably visit an absorbing, disease-free state at
sufficiently long times due to the finite size of the networks where epidemics
evolves. In the present work, we compare different quasistationary (QS)
simulation methods where the absorbing states are suitably handled and the
thermodynamical limit of the original dynamics can be achieved. We analyze the
standard QS (SQS) method, where the sampling is constrained to active
configurations, the reflecting boundary condition (RBC), where the dynamics
returns to the pre-absorbing configuration, and hub reactivation (HR), where the
most connected vertex of the network is reactivated after a visit to an
absorbing state. We apply the methods to  the contact process (CP) and
susceptible-infected-susceptible (SIS) models on regular and scale free
networks. The investigated methods yield the same epidemic threshold for both
models.  For CP, that undergoes a standard collective phase transition, the
methods are equivalent. For SIS, whose phase transition is ruled by the hub
mutual reactivation, the SQS and HR methods are able to capture localized
epidemic phases while RBC is not. We also apply the auto-correlation time as a
tool to characterize the phase transition and observe that this analysis
provides the same finite-size scaling exponents for the critical relaxation time
for the investigated methods. Finally, we verify the equivalence between RBC
method and a weak external field for epidemics on networks.

\end{abstract}

\author{Renan S. Sander}

\author{Guilherme S. Costa}

\author{Silvio C. Ferreira}

\affiliation{Departamento de F\'{\i}sica, Universidade Federal de
  Vi\c{c}osa, 36570-900, Vi\c{c}osa, Brazil} 


\maketitle

\section{Introduction}
Nonequilibrium phase transitions are observed in many physical, biological,
social, and chemical systems~\cite{Henkel,hinrichsen,odor04}, to mention just a
few examples. A fundamental class of nonequilibrium phenomena is the
absorbing-state phase transition (APT)~\cite{Henkel,hinrichsen,marro1999npt}
that occurs when the system has accessible states where the dynamics remains
permanently stuck, strongly violating detailed balance and thus implying the
absence of equilibrium counterparts. One of the characteristic features of APTs
is the interplay between the propagation and annihilation of  particles that can
represent hosts for an infectious disease, microorganisms, molecules in
catalytic reactions, etc. When the ratio between propagation and annihilation is
below  a threshold, the density $\rho$ of active particles vanishes and an APT
takes place.

A fundamental model that exhibits an APT is the contact process
(CP)~\cite{harris74}, where infected individuals lying on a substrate
(originally a lattice~\cite{harris74} but it was later extended to
networks~\cite{RomuPRL2006}) can spontaneously be cured or infect one of their
nearest neighbors. From the epidemiological viewpoint, the
susceptible-infected-susceptible (SIS) model is canonical~\cite{pv01a}. In SIS,
the cure is the same as in CP but the infection is transmitted  at a rate
$\lambda$ to each healthy neighbor of an infected vertex while in CP the
infection rate is divided by the vertex degree. On lattices, SIS and the CP have
the same critical behavior belonging to the directed percolation (DP)
universality class~\cite{marro1999npt}. However, the nature of the phase
transition of these models are  different on complex networks with power-law (PL)
degree distributions given by $P(k) \sim k^{-\gamma}$, where $P(k)$ is the
probability that a randomly chosen vertex of the network connects to other $k$
vertices. While CP exhibits a  transition at a finite threshold that involves
the collective activation of the entire network~\cite{Ferreira16}, the
transition of the SIS model occurs at a threshold that goes to zero in the
thermodynamical limit and is triggered by the mutual reactivation of hubs (
vertices ith very large degree)~\cite{Ferreira16,boguna2013}.

Many dynamical processes exhibiting distinct types of  APTs have been
intensively studied on lattices~\cite{Henkel, hinrichsen, odor04, marro1999npt}.
However, the behavior of this kind of process is strongly affected by the
structure, in the form of a network, that connects the system
components and mediate their interactions~\cite{dorogovtsevRMP08,
RMPreview2014}. The former investigations of an epidemic spreading on complex
networks~\cite{pv01a, RomuPRE01,May01,Newman02}, which later revealed many
remarkably properties and puzzling outcomes~\cite{RomuPRL2010, chatterjee2009,
Lee2013, Mata15, Goltsev12, Odor13}, were subsequently followed by a diversified
analysis of other kinds of APTs on networks~\cite{Goh03,Bancal10,Lee04,da-yin10,
Sander13,RomuPRL2006, HongPRL2007, RomuPRL2008, Ferreira_quenched}.

The large amount of gathered data regarding real networks~\cite{newman2010}, the
recent advances on complex network theory and the increased computational power
are among the factors that leveraged the modern research on networks and
dynamical processes taking place on them. Even for simple dynamical processes
most of the exact results are bounds~\cite{RMPreview2014} and analytical
approximations based on mean-field theories are generically used to
quantitatively predict the behavior of such processes on
networks~\cite{dorogovtsevRMP08,RMPreview2014}. Dynamical correlations play an
important role irrespective of the infinite dimensionality of
networks~\cite{mata2013pair,mata2014,Gleeson} and  mean-field theories are
approximations that call for complementary analysis based on computer
simulations. For example, the existence of localized epidemic
phases~\cite{Lee2013,Mata15,Buono,odor2014localization} and the slow subcritical
dynamics due to rare (locally supercritical)
regions~\cite{MunozPRL2010,Odor13,Buono,Cota16}   on networks with PL
degree distribution  have recently been discussed grounded on simulations.
Notice that localization is related to  metastability at  subcritical
phases~\cite{MunozPRL2010} and can play an important role in the quasi stationarity
of the dynamical process.

A major hurdle in the simulations of APTs is that they are unavoidably performed
in finite size systems  and for sufficiently long times an absorbing state is
always visited due to the finite number of accessible configurations  implying
that the unique real stationary state  is an absorbing one~\cite{marro1999npt}.
Moreover, standard dynamic methods for simulations of APTs as, e.g., spreading
or decay simulations~\cite{marro1999npt}, are not effective in complex networks due to
the small-world property~\cite{newman2010} that makes the dynamics to probe the
network finiteness very quickly. An alternative is to consider the so-called
{quasistationary} (QS) state where the original dynamics is perturbed to skip
the absorbing state in a such way that this perturbation becomes irrelevant 
in  the infinite size limit for all intensive quantities of interest and a
finite size scaling (FSS) technique provides the correct thermodynamical limit
as well as the critical exponents associated with the APT~\cite{Henkel,marro1999npt}.

The standard QS (SQS) method consists in performing averages only
over the samples that did not visit an absorbing state~\cite{marro1999npt}.
Also, a reflecting boundary condition (RBC), where the evolution 
returns to the pre-absorbing state when it visits an absorbing one, can be
easily applied~\cite{dickman_tania}.  Another approach\footnote{There exist
other important methods to deal with absorbing states in finite size systems,
e.g. those conserving the number of particles~\cite{Tome01}.} is to include an
external field (EF)~\cite{Henkel,lubeck2003,gunnar}, conjugated to the order
parameter, that continuously  and non correlatively introduces activity.
If the field is sufficiently  weak it becomes equivalent to the RBC
method~\cite{gunnar}, which in turn, provide the same critical points and FSS of
the critical density of the SQS method in lattice systems~\cite{gunnar,Mancebo2005}.

The SQS method has been used to investigate the APT of the CP model on networks
with a PL degree distribution showing very good agreement with the critical
exponents obtained by a heterogeneous mean-field  (HMF)
theory~\cite{Ferreira_quenched,mata2014}. This method was also applied to
numerically determine the epidemic threshold of the SIS model as a function of
the network size~\cite{Ferreira12,Mata15}. Besides a vanishing
threshold~\cite{Ferreira12}, the SQS method was applied to investigate multiple
transitions~\cite{Mata15} and localization~\cite{Cota16}  of the SIS model on
networks with a PL degree distribution. The EF method with a finite source was
also applied to the SIS model~\cite{Mieghem12}. In Ref.~\cite{Lee2013}, a
simulation strategy was adopted wherein the most connected vertex of the network is never
cured. However, a systematic comparison among different methods
remains lacking.

In this work, we compare the outcomes of  RBC and SQS methods in the APT of the
CP and SIS models in regular and PL networks. We apply the integrated
autocorrelation time of QS time series as a measure of the characteristic
relaxation time instead of the usual average time between two consecutive visits
to be  the absorbing state because the latter does not represent the relaxation
time in RBC method. For heterogeneous networks, we also investigate an
alternative version of the RBC method, called hub reactivation (HR), where the
dynamics always restart in the most connected vertex of the network. This method
resembles the strategy used in Ref.~\cite{Lee2013}. Finally, we extend the
results of Ref.~\cite{gunnar} to networked substrates, confirming that the weak
limit of EF (WEF) method is equivalent to RBC.

In summary, we observe that the SQS, RBC, and HR methods provide the same
threshold and critical exponents for CP on complex networks and regular lattices.
For the SIS model on networks with a PL degree
distribution~\cite{barratbook}, SQS and HR methods differ from the RBC method in
some relevant aspects. For instance, the first and second ones capture activated
configurations which are highly localized around the most connected vertex of
the network, while the last one does not.

The sequence of the paper is organized as follows. The theoretical background
for the different QS methods is presented in Sec.~\ref{sec:qs}.  In
Sec.~\ref{sec:numerical}, we describe CP and SIS models with their computer
implementations. Numerical methods to characterize the critical point including
the integrated autocorrelation time are presented in Sec.~\ref{sec:method}. The results
along with the associated discussions are in Sec.~\ref{sec:results}
and we summarize our findings in Sec.~\ref{sec:conclu}.

\section{Quasistationary states}
\label{sec:qs}
The terminology QS is commonly attributed to averages restricted to samples that
did not visit an absorbing state of the original dynamics~\cite{marro1999npt}.
Here, this term is used in a more general context, in which the dynamical
process is perturbed to prevent the system from getting trapped into absorbing states
but assuring that intensive QS quantities converge to the stationary ones in the
thermodynamical limit.

In order to comprehend some basic concepts necessary to develop the QS methods,
let us consider  a one-step process $X_{t}$, in which only transitions
$n\rightarrow n\pm 1$ are permitted~\cite{vankampen},  with states labeled by
$n=0,1,2,\ldots$,  and the state $n=0$ is absorbing. Now, lets $X_{t}^{*}$ be a
similar process where the unique difference is that the state $n=0$ is not
absorbing anymore. For  $n>0$, the evolution of $X_{t}^{*}$ is the same as
$X_{t}$ unless, at most, by some perturbation in the rules that must be
negligible in the active phase in the thermodynamical limit. The process
$X^*_{t}$ returns to a state $n>0$ after some time, whenever the system
visits an absorbing state.

Let $P_n$ and $P_n^*$ be the probabilities to be in the state $n$ in the original 
and modified dynamics, respectively. The master equation~\cite{vankampen} for the
original process reads as
\begin{equation}
\frac{dP_n}{dt} = \sum_{m}w_{nm}P_m-\sum_{m}w_{mn}P_n,
\label{eq:ME}
\end{equation}
where $w_{nm}$ is  the transition rate from  $m$ to $n$. Introducing a
perturbative source of activity to remove the absorbing state, the new dynamical
equation\footnote{For the modified dynamics it is not necessarily a master
	equation since non-linear terms are allowed.}
becomes
\begin{equation}
\frac{dP_n^*}{dt} = \sum_{m}w_{nm}P^*_m-\sum_{m}w_{mn}P^*_n + F(P_0^*,P_1^*,\ldots)
\label{eq:ME2}
\end{equation}
where the functional $F$ will depend on the particular QS method. The basic QS
quantities of interest can extracted from the QS distribution $\bar{P}_{n}$
yielded by the  stationary solution of Eq.~\eqref{eq:ME2}.

\subsection{Standard QS method}
\label{sec:sqs}

This method consists in performing averages only over samples that did not
visit an absorbing state. The QS distribution of the
original dynamics is given by 
\begin{equation}
\bar{P}_n=\lim\limits_{t\rightarrow\infty}  \frac{P_n(t)}{P_s(t)},~~~n>0
\label{eq:QSdef}
\end{equation}
where
\begin{equation}
P_s(t)=\sum_{n\geqslant 1} P_n(t)=1-P_0(t)
\label{eq:Ps}
\end{equation}
is the probability that the epidemics is active  at time $t$. In practice, this
strategy is troublesome since critical and subcritical simulations in finite
size get constantly trapped into absorbing states resulting in short and noisy
intervals of stationary data. de~Oliveira and Dickman~\cite{Mancebo2005}
proposed a clever strategy to circumvent these problems: Every time the dynamics
visits an absorbing state, the system jumps to an active configuration selected according
to the QS probability. The source term in Eq.~\eqref{eq:ME} is  thus given by
$F=w_0 P_{n}^{*}$~\cite{Mancebo2005}, where $w_0=\sum_{m\ne 0}w_{0,m}$ is the total
rate of entering into the absorbing state. One can verify that the
stationary solution of Eq.~\eqref{eq:ME2} with this source term corresponds to
the QS solution of Eq.~\eqref{eq:ME}; see Ref.~\cite{Mancebo2005} for details.

Differentiating Eq.~\eqref{eq:Ps} and assuming that the QS regime
exists as $t\rightarrow\infty$ we find that
\begin{equation}
\frac{dP_s}{dt} = -\frac{dP_0}{dt} = -\bar{P}_1 P_s(t),
\label{eq:dPsdt}
\end{equation}
where we used the master equation~\eqref{eq:ME} for $n=0$ and $P_n = \bar{P}_n
P_s$ for $n>0$. The solution is $P_s\sim \exp (-t/\tau_a)$, where
\begin{equation}
\tau_a=\frac{1}{\bar{P}_{1}},
\label{eq:tau}
\end{equation}
is the typical relaxation time to decay from an active QS state to 
an absorbing configuration that also corresponds to the time between two
consecutive visits to an absorbing state during the QS regime.

The difficulty to translate this theoretical analysis into a simulation scheme
is that  we have no prior knowledge of the QS probability distribution.
Computationally, it can be  done by constructing and constantly updating a list
with $M$ configurations visited along the simulation.  This list is used  to
randomly select the new state after a visit to an absorbing state. To grant
convergence to the QS state, the list, which is finite,  is constantly updated
by substituting a randomly selected element of the list  by the current system
configuration with probability per unit of time $p_{r}$. The convergence of this
method to  the standard QS state, defined by Eq.~\eqref{eq:QSdef}, was recently addressed~\cite{blanchet}.

\subsection{Reflecting boundary condition method}
\label{sec:rbc}
The absorbing phase can be avoided by bringing back the system to the
configuration that it was immediately before the visit to the absorbing state,
representing a reflecting boundary condition~\cite{dickman_tania}. We permit
states with $n=0$ being visited but it returns to the previous active state with
rate 1. The source term in Eq.~\eqref{eq:ME2} becomes
$F=(\delta_{1,n}-\delta_{0,n})P^*_0$. The QS distribution is given by 
\begin{equation}
\bar{P}_n=\lim\limits_{t\rightarrow\infty} P^*_n. 
\end{equation}
For $n=0$, Eq.~\eqref{eq:ME2} simplifies to
\begin{equation} 
\frac{d P_0^*}{dt} = P_1^*-P_0^*. 
\end{equation} 
In the QS regime we have $\bar{P}_1=\bar{P}_0$. To calculate the time between
two visits to the absorbing state, the epidemic lifespan, consider a discrete
dynamics of time step $\tau_0=1$, the mean time that the system lasts in the
absorbing state. The system is or is not in a state $n=0$ with probabilities
$\bar{P}_0$ and $1-\bar{P}_0$, respectively. If the system is active in step 
$s=0$, the probability that  it stays active for $s$ steps corresponds to stay
in a state $n>0$  for $s-1$ steps and returns to $n=0$ at step $s$, which  is given by
$Q_s=(1-\bar{P}_0)^ {s-1}\bar{P}_0$. The average time between two visits to the
absorbing state is
\begin{equation}
\tau_a = \sum\limits_{s=0}^{\infty} \tau_0sQ_s = \frac{\tau_0}{\bar{P}_0}=\frac{1}{\bar{P}_1},
\label{eq:tauarbc}
\end{equation}
recovering Eq.~\eqref{eq:tau}. Notice that this time differs from that of
the SQS method since it is the average time to visit the absorbing state
starting from a single infected vertex and not from a arbitrary QS state, as discussed in subsection II.A. Thus its scaling properties are
different. Lets us consider a spreading process on a lattice of dimension $d$.   The
probability that the dynamics is active at time $t$ starting with a single
infected vertex is $P_s$ and, at the critical  point, this quantity scales as
$P_s\sim t^{-\delta}\exp(-t/t_s)$~\cite{marro1999npt} where the finite size of
the lattice is probed at a characteristic time $t_s\sim
N^{z^*}$~\cite{marro1999npt}, being $N=L^d$ the number of sites\footnote{It is
	more convenient to consider the number of sites rather than the system length
	since we aim at networks.} of the lattice of length $L$. So, the average time
between two visits to the absorbing state is given by
\begin{equation}
\tau_a = \int_0^{\infty}t\frac{dP_0}{dt}dt=\int_0^{\infty}P_s dt\sim N^{z^*(1-\delta)}.
\label{eq:tauaRBC}
\end{equation}
for $d$ smaller than the upper critical dimension $d_c=4$ for which
$\delta<1$~\cite{marro1999npt}. The last integral is obtained with
Eq.~\eqref{eq:Ps} and an integration by parts and was evaluated using the saddle-point method. Above the upper critical dimension, corresponding to the
mean-field level that we are mainly interested in, we have $\delta=1$ and
$t_s\sim N^{1/2}$~\cite{Janssen2005,dickman2002quasi}. Evaluating the integral
for $\delta=1$, we find $\tau_a$ increasing logarithmically with  the system
size.

\subsection{Hub reactivation method}
\label{sec:hr}

In the case of heterogeneous networks, we also investigate an alternative to the RBC
method where, after visiting the absorbing state, the infection always restarts
in the most connected vertex of the network or in one of them if there are
multiple. This strategy favors the onset of outbreaks since hubs are usually
prone to spread the activity. The motivation of this method is the existence of
localized active phases around the hubs in SIS-like dynamics~\cite{Mata15},
which are suppressed by the RBC method. In Sec.~\ref{sec:results}, this point
will be made clear.

\subsection{External field method}
\label{sec:ef}
In this method, the system is coupled to a uniform external field that spontaneously
creates activity at a rate $f$ chosen to vanish as $N \rightarrow
\infty$~\cite{gunnar}. The source term in Eq.~\eqref{eq:ME2} is  $f(N-n+1)
P^*_{n-1}-f(N-n)P^*_n$, $n>0$, that  accounts for the incomes and outcomes of
$P_n^*$ due to the spontaneous creation. For the absorbing state we have
\begin{equation} 
\frac{d P_0^*}{dt} = P_1^*-fNP_0^*,
\end{equation}
implying  $\bar{P}_1^*=fN\bar{P}_0^*$ for $t\rightarrow\infty$. Using Eq.~\eqref{eq:tauarbc} we have that
$\tau_a=\tau_0/\bar{P}_0^*=1/\bar{P}_1^*$ where $\tau_0=1/(fN)$ was used. 

The critical density produced by a small external field $f$ scales as
$\bar{\rho}_\mathrm{ex} \sim f^{1/\delta_{h}}$, where $\delta_h$ is a critical
exponent~\cite{marro1999npt}. The QS density scales as $\bar{\rho}_\mathrm{qs}
\sim N^{-\beta^*}$. Imposing that the critical density of particles produced by
the external field is negligible  when compared with the QS density and assuming
$f\sim N^{-\alpha}$, we have that the condition $\bar{\rho}_\mathrm{ex}\ll
\bar{\rho}_\mathrm{qs}$ is satisfied for $\alpha > \beta^*\delta_{h}\ge 1$, the
last inequality is verified considering the critical exponents
$\beta^*=(0.252,0.398,0.464,1/2)$ and $\delta_h=(9.23,3.72,2.52,2)$ for DP in
$d=1,2, 3$ and $d\ge 4$, respectively~\cite{Henkel}. The total creation in
absorbing states goes to zero in the thermodynamical limit and the dynamics gets
trapped for diverging times into the absorbing states. Computationally it is not
a problem since it is implemented as a time step $\Delta t = 1/fN$; see
Subsection.~\ref{sec:algor}. Hence, we must construct the QS distributions using
only the  non-absorbing part of the simulations as
\begin{equation}
\bar{P}_{n}\equiv \lim\limits_{t\rightarrow\infty}\frac{P_n^*}{\sum_{m>0}P_m^*}
\end{equation}
and $\bar{P}_0\equiv 0$. 

\section{Models and algorithms}
\label{sec:numerical}
\subsection{Models}
\label{sec:models}

The CP model was originally proposed as a simple model for epidemic
propagation on a lattice~\cite{harris74}. For a more general case,  we consider
a connected graph with $N$ vertices and a quenched (not changing in time)
connection structure. Each vertex  $i$ of the network has $k_i$ edges connecting
to nearest neighbors and can be either infected ($\sigma_{i}=1$) or susceptible
($\sigma_{i}=0$).  Infected vertices spread activity through the
network by direct susceptible  contacts. An infected vertex $i$
transmits the infection to each of its susceptible nearest-neighbors with a rate
$\lambda/k_i$, in which $\lambda$ is the control parameter. In turn, infected
vertices become spontaneously susceptible with rate 1 fixing the time unit. In
SIS model, the infection rate is $\lambda$ for each edge connecting infected and
susceptible vertices irrespective of their degrees while the healing process is
the same of CP. So, one can easily realize that the state $\sigma_{i}=0$ for
all vertices is absorbing in both models. The limit between persistence
and extinction of activity is delimited by a critical value $\lambda_{c}$ of the
control parameter. The density of infected vertices at the stationary state,
$\bar{\rho}$, is the order parameter that is null in the absorbing phase
$\lambda<\lambda_{c}$ and  reaches a steady value for $\lambda>\lambda_{c}$.

Contrasting with homogeneous graphs, SIS and CP are very different for
heterogeneous networks. Distinct analytical approaches were devised to determine
an expression for $\lambda_{c}$ and the respective critical exponents in random
networks with PL degree distributions, for both SIS~\cite{pv01a,
boguna2013, Goltsev12, RomuPRL2010, Wang03, Mieghem11, Mieghem12} and CP
\cite{RomuPRL2006, RomuPRL2008, bogunaPRE2009, NohPRE2009, HongPRL2007,
Ferreira_annealed, mata2014}. The central results are that, in the
thermodynamical limit,  CP has a finite threshold and undergoes an APT with well
defined FSS exponents. For infinite size PL networks, the SIS model does
not exhibit an APT such that an endemic active phase is observed for any finite
infection rate $\lambda>0$. However, for finite sizes the SIS effectively has a
threshold~\cite{RomuPRL2010,Ferreira12,Lee2013,mata2013pair,boguna2013,Mata15}
and the APT, including FSS exponents, can also be investigated numerically.

\subsection{Algorithms}
\label{sec:algor}

For the SQS, RBC, and HR methods with $n>0$, the simulations of the SIS dynamics
were performed according the adapted Gillespie algorithm presented in
Refs.~\cite{Ferreira12,Mata15}. At each time step, the number of infected
vertices $N_{i}$ and the sum of the degrees of these vertices $N_{n}$ are
computed. With a probability $N_{i}/(N_{i}+\lambda N_{n})$ a randomly chosen
infected node becomes healthy. With the complementary probability $\lambda
N_{n}/(N_{i}+\lambda N_{n})$, an infected vertex is chosen proportionally to its
degree and one of its links is chosen with equal chance. If the selected link
points to a susceptible vertex it becomes infected; otherwise the simulation
continues. The total number of infected nodes and the number of links emanating from them
are updated, time is incremented by $t \rightarrow t + \Delta t$,
where\footnote{In the Gillespie algorithm, the time increments are drawn from an
	exponential distribution with average $\Delta t$. For long QS averaging, the
	results are independent of this step.} $\Delta t=1/(N_{i}+\lambda N_{n})$, and
the entire process is iterated. If the system visits the absorbing state,  the
rules described in Subsecs~\ref{sec:sqs}, \ref{sec:rbc}, or \ref{sec:hr} are
applied, returning to an active state. It is worth noting that this algorithm
is statistically exact in the same sense as the Gillespie algorithm.
 
For the EF method, a third possible event is defined: The spontaneous infection
of a randomly chosen site with rate $f$. In this case, a chosen infected node
becomes healthy with probability $N_{i}/(N_{i}+\lambda N_{n}+fN)$, the
transmission of the infection is tried with probability $\lambda
N_{n}/(N_{i}+\lambda N_{n}+fN)$, while spontaneous creation at a randomly chosen
vertex is tried with the complementary probability $fN/(N_{i}+\lambda
N_{n}+fN)$. If the chosen vertex is susceptible it becomes infected otherwise
the simulation goes on. The time increment is $\Delta t=1/(N_{i}+\lambda
N_{n}+fN)$.

The simulations of the CP were done using the standard
procedure~\cite{marro1999npt}. At each time step, an infected vertex is chosen
with equal chance and  attempts to infect one of its nearest-neighbors, randomly
selected, with probability $\lambda/(1+\lambda)$. Annihilation
of a randomly chosen infected vertex occurs with the complementary probability
$1/(1+\lambda)$. Time is incremented by $\Delta t = 1/(N_{i}+\lambda N_{i})$ and
the whole process is iterated. For the EF method, the probabilities are
$fN/(N_{i}+\lambda N_{i}+fN)$, $\lambda N_{i}/(N_{i}+\lambda N_{i}+fN)$, and
$N_{i}/(N_{i}+\lambda N_{i}+fN)$ for spontaneous infection, catalytic infection
and spontaneous healing, respectively, and the time step is $\Delta t =
1/(N_{i}+\lambda N_{i}+fN)$.

The relevant QS quantities are calculated during an averaging time
$t_\mathrm{av}$, after a relaxation time $t_\mathrm{rlx}$. The QS probability
$\bar{P}_{n}$ is computed during the interval $t_\mathrm{av}$,  such that each
active configuration contributes to the QS distribution with a weight
proportional to its lifetime $\Delta t$. In networks, we used typically
$t_\mathrm{av}=10^{8}$, $t_\mathrm{rlx}=10^{6}$, and $p_{r}=10^{-2}$, the last
one in SQS method. For lattices, $t_\mathrm{rlx}>10^7$ was used. For very
supercritical simulations shorter times are used.

\section{Numerical characterization of the critical point}
\label{sec:method}

\subsection{Susceptibility and epidemic lifespan.}
\label{sec:sus}

\begin{figure}[ht]
\centering
\includegraphics[width=0.9\linewidth]{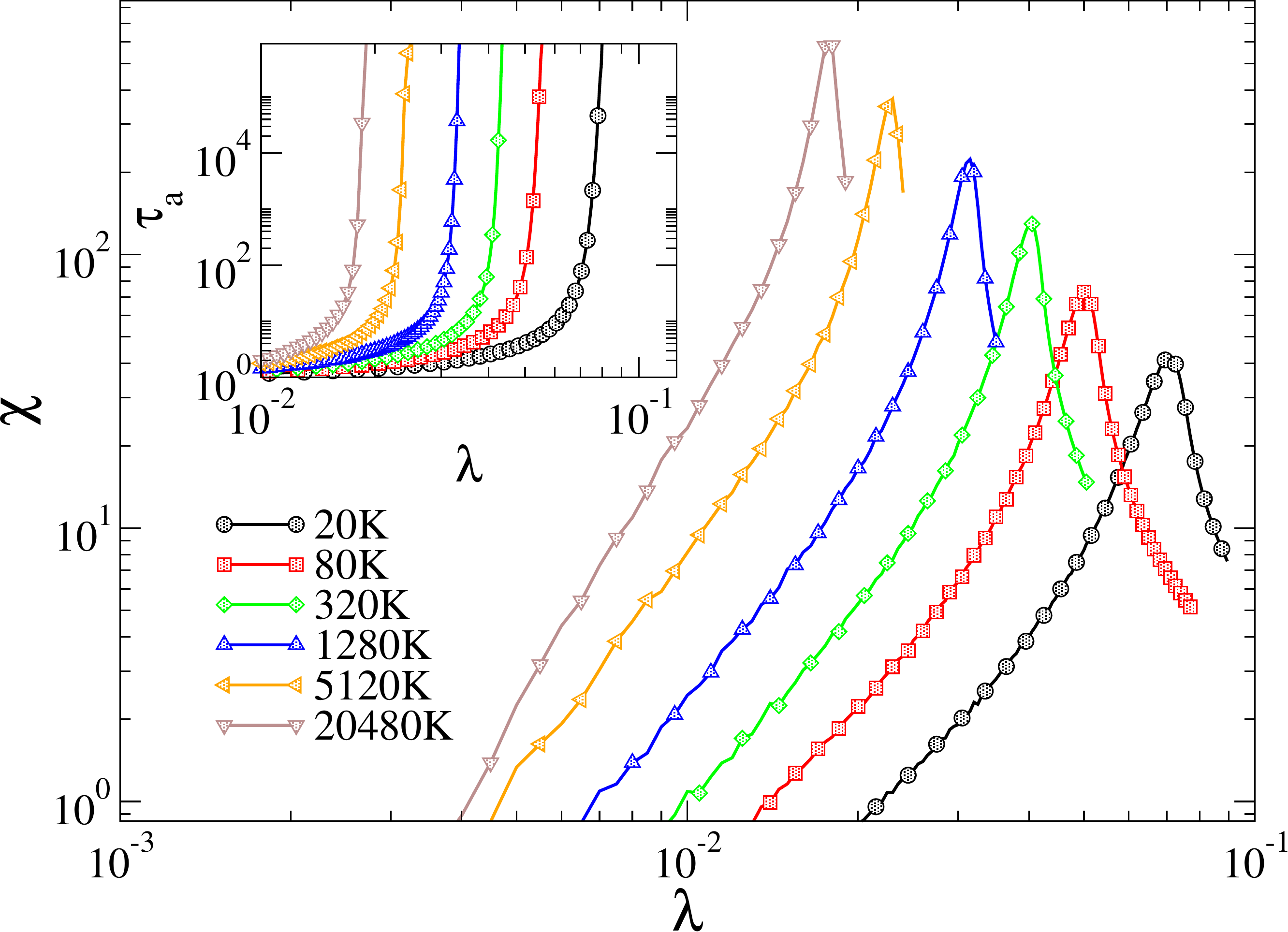}
 \caption{Main panel: Susceptibility vs infection rate for the SIS model on UCM
 networks~\cite{Catanzaro05} with a PL degree distribution  with
 $\gamma=2.7$ and different sizes using the SQS method. Inset: Lifespan vs
 infection rate. The network sizes are indicated in the legend.}
 \label{fig.suscep_sis_qe_270}
\end{figure}

The critical behavior of non-equilibrium processes presents diverging
correlation length and time. The former makes little sense for complex networks
because of the small-world property stating that average distance  between two
vertices increases logarithmically or slower with the network size~\cite{newman2010}. However, the
concept of diverging temporal correlations can be applied to identify and
characterize the APT~\cite{Ferreira12,mata2014,Mata15,Lee2013,deArruda,Shu}. The
susceptibility defined as~\cite{Ferreira12}
\begin{equation}
\label{eq.suscep}
 \chi = N \frac{\langle \rho^{2} \rangle - \langle \rho \rangle^{2}}{\langle \rho \rangle}
\end{equation}
 exhibits a peak that diverges as size increases at the transition point
for SIS on  uncorrelated networks with degree exponent
$\gamma<3$~\cite{Ferreira12,mata2013pair,Lee2013} and CP on complex networks in
general~\cite{mata2014,Ronan2013}. This peak will be used to estimate the
effective, size-dependent thresholds $\lambda_{p}$.
Figure~\ref{fig.suscep_sis_qe_270} shows the susceptibility $\chi(\lambda)$ and
the lifespan $\tau_a(\lambda)$ (inset) for different network sizes $N$ obtained
with the SQS method. Note that for all curves shown, $\tau_a$ diverges
approximately at the same position $\lambda_{p}$ where the peaks of the
susceptibility $\chi$ curves are located. Using the lifespan $\tau_a$ as an
order parameter is not possible due to its divergence in the active phase.
Hence, we consider the integrated autocorrelation time in the next subsection. In the case of
multiple peaks in the susceptibility curves~\cite{Ferreira12}, the epidemic
threshold is taken as the one occurring nearest to the lifespan
divergence~\cite{Mata15}.

\subsection{Autocorrelation time}
\label{sec:autocorr}
\begin{figure}[th]
\centering
\includegraphics[width=0.9\linewidth]{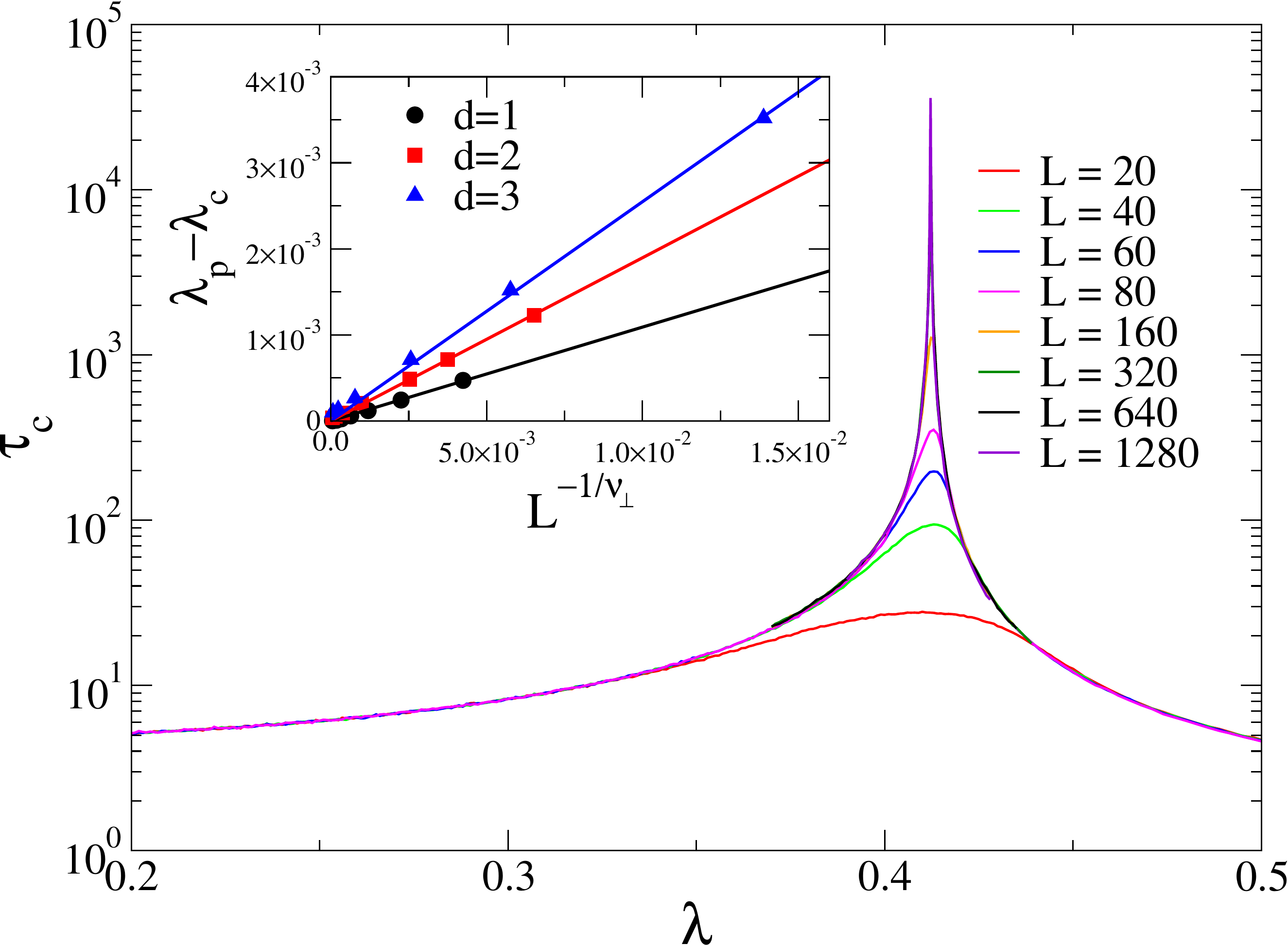}
\caption{Autocorrelation time against infection rate for SIS model on square
lattices of different sizes being the sharper the larger the size. The simulation method is SQS. Inset shows the
FSS of the epidemic threshold for $d=1,2,3$.}
\label{fig:taucSIS2d}
\end{figure}
The autocorrelation time of a series is defined as~\cite{NR2007} 
\begin{equation}
	\tau_c=\frac{1}{2}\sum\limits_{s=0}^{t_\mathrm{av}} C(s),
\end{equation}
where $C(s)$ is the autocorrelation function given by
\begin{equation}
C(s)=\frac{\lrangle{\tilde{\rho}(s'+s)\tilde{\rho}(s')}}{\lrangle{\tilde{\rho}^2}},
\end{equation} 
and $\tilde{\rho}(s)=\rho(s)-\lrangle{\rho}$ is a time series built recording
the density in the QS regime between time intervals $\Delta t=1$. Here, brackets
represent time averaging over $s'$ at the QS regime. To prevent spurious behavior in the subcritical phase
with the SQS method, the time series must be rid of big gaps. So, every time a
new configuration is randomly selected  to replace the absorbing state, we
discard an interval of the time series  in a such a way that the number of
infected vertices before and after the replacement differs at most by $\Delta n
=\pm 1$. For RBC, HR and EF methods this problem does not exist. Large time series
with at least $10^7$ points were used to calculate $\tau_c$ near and below the
epidemic thresholds. The autocorrelation function of long time series is
efficiently computed as the inverse Fourier transform of the power spectrum of
the series~\cite{NR2007}.

To validate the method, we performed simulations of the critical SIS model on
hypercubic lattices of dimensions $d=1$, 2, and 3 with periodic boundary
conditions. Figure~\ref{fig:taucSIS2d} shows the autocorrelation time against
infection rate for SIS model in square lattices of different sizes exhibiting a
pronounced diverging peak as the lattice size increases. The position of the
peak $\lambda_p$ converges to the expected SIS threshold\footnote{In regular
graphs where all vertices have the same number of connections $k$ the SIS and CP
thresholds are related by $\lambda_{c}^\mathrm{SIS}=\lambda_c^\mathrm{CP}/k$.
Using the thresholds known for CP in hypercubic lattices~\cite{Henkel} we have
$\lambda_c^\mathrm{SIS}=1.648924$, 0.41219, and 0.21948 for $d=1$, 2, and
3, respectively.} $\lambda_c$ in $d=1$, 2, and 3 following the standard
FSS~\cite{Henkel}
\begin{equation}
\lambda_p-\lambda_c\sim L^{-1/\nu_\perp},
\end{equation}
where $\nu_\perp$ is the critical exponent associated with the divergence of 
the correlation length; see the inset of Fig.~\ref{fig:taucSIS2d}.

\section{Comparison of the SQS and RBC  methods in regular graphs}

\begin{figure}[th]
\centering
\includegraphics[width=0.9\linewidth]{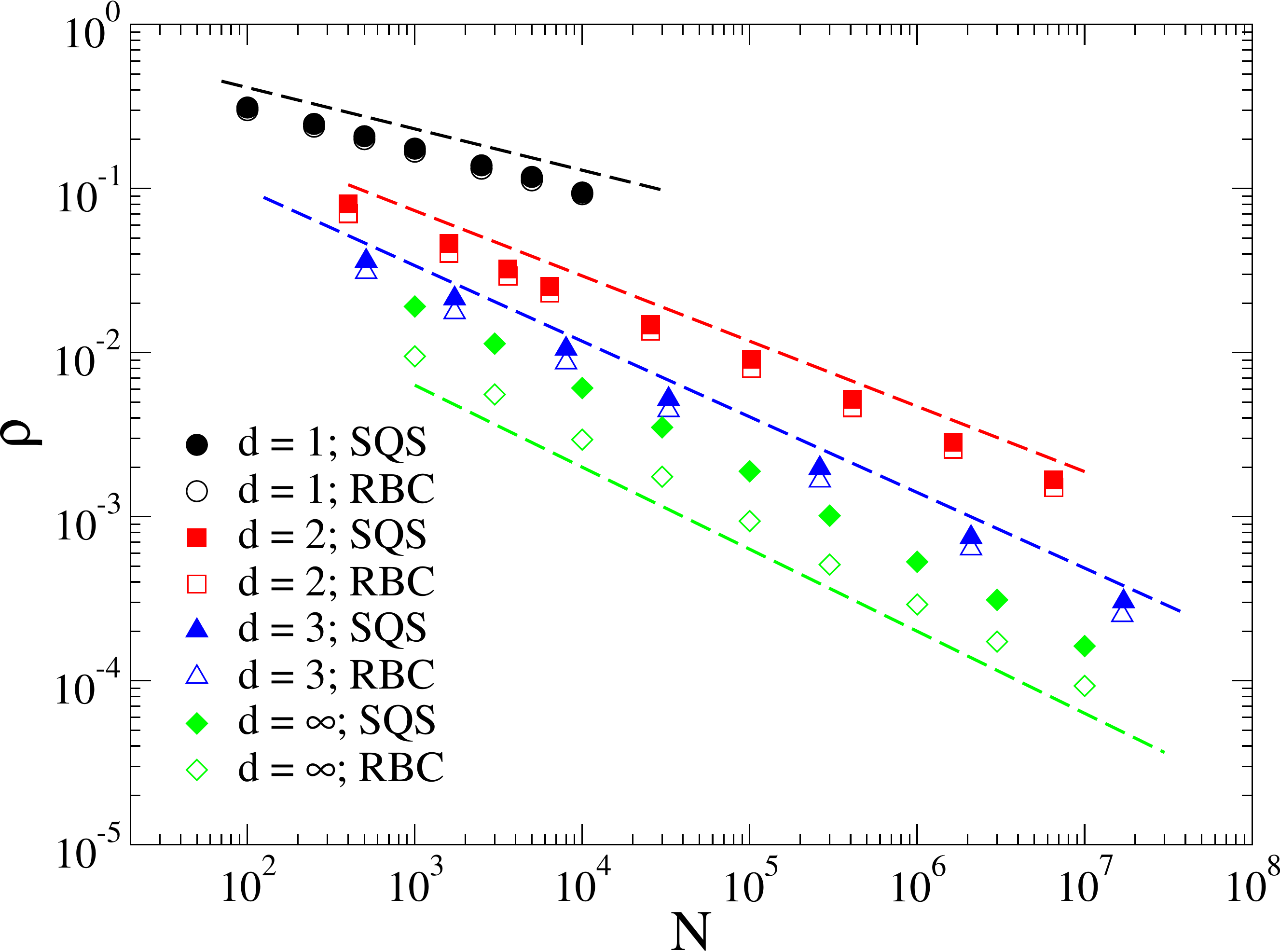}
\caption{QS density against number of vertices for critical SIS on lattices
of dimensions $d=1$, 2, 3 and RRN ($d=\infty$) using SQS  and RBC methods.
Dashed lines represents the scaling  exponents of the DP class~\cite{Henkel}.}
\label{fig:rhoSISlattice}
\end{figure}

Figure~\ref{fig:rhoSISlattice} shows the QS density for critical SIS against the
number of vertices $N$ ($N=L^d$ for lattices) in $d$-dimensional hypercubic
lattices and in random regular networks (RRN)~\cite{Ronan2013}. In the last one,
all vertices have the same degree $k=3$  and connections are
random~\cite{Ferreira12} implying that RRN corresponds to an infinite dimension.
Power-law decays in the form $\bar{\rho}\sim N^{-\beta^*}$ with the expected
exponents $\beta^*=0.2521$, $0.3978$, $0.4640$,  and $1/2$ for DP class in
dimensions $d=1,~2,~3,$~\cite{Henkel} and $\infty$~\cite{Janssen2005} are
observed in both SQS and RBC methods.
The different  QS methods correspond to distinct strengths of perturbation of 
the system. Thus the critical quantities do not have to be identical but
only present  the same scaling to grant equivalence among methods.

\begin{figure}[th]
\centering
\includegraphics[width=0.9\linewidth]{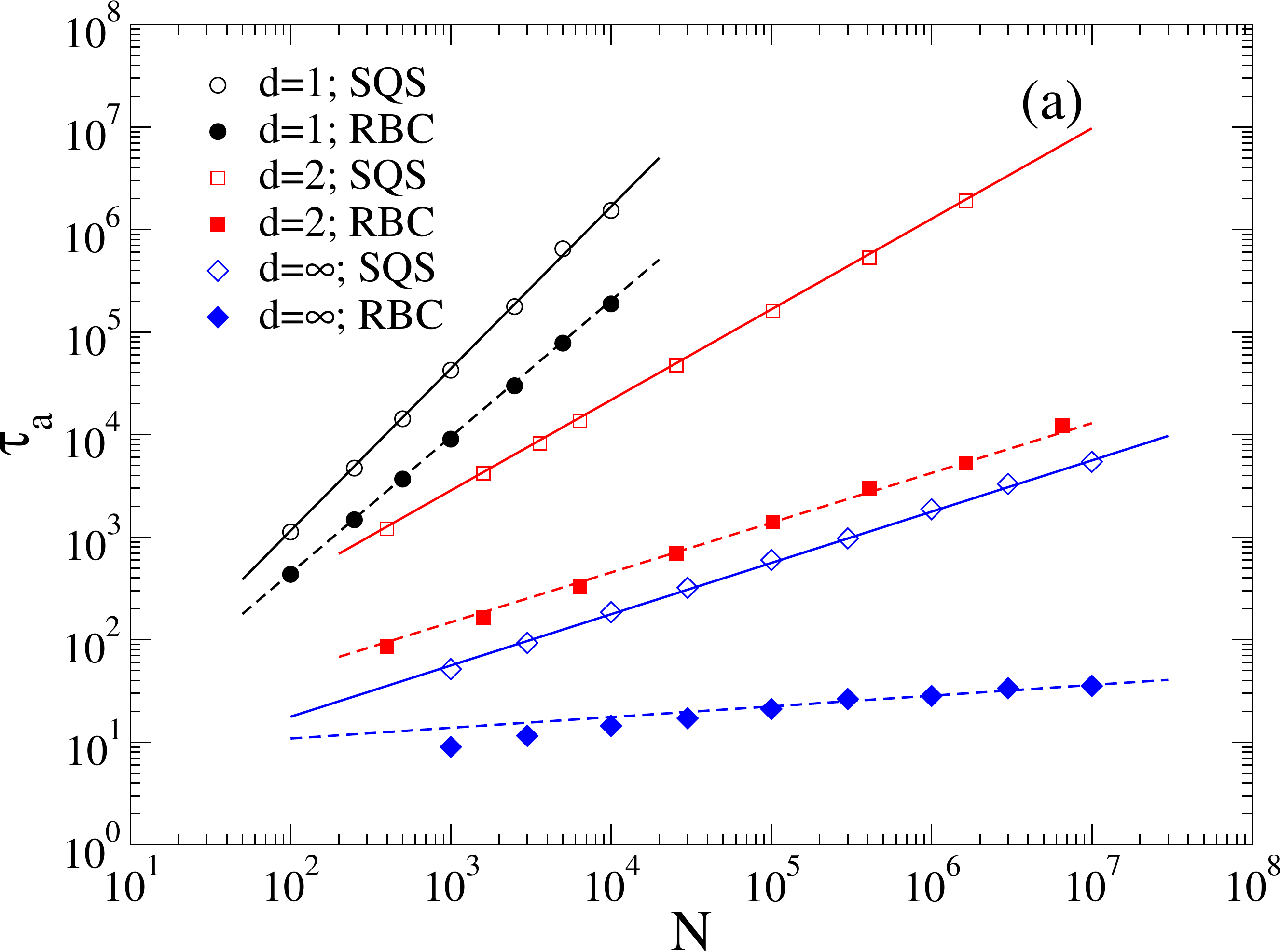}\\
\includegraphics[width=0.9\linewidth]{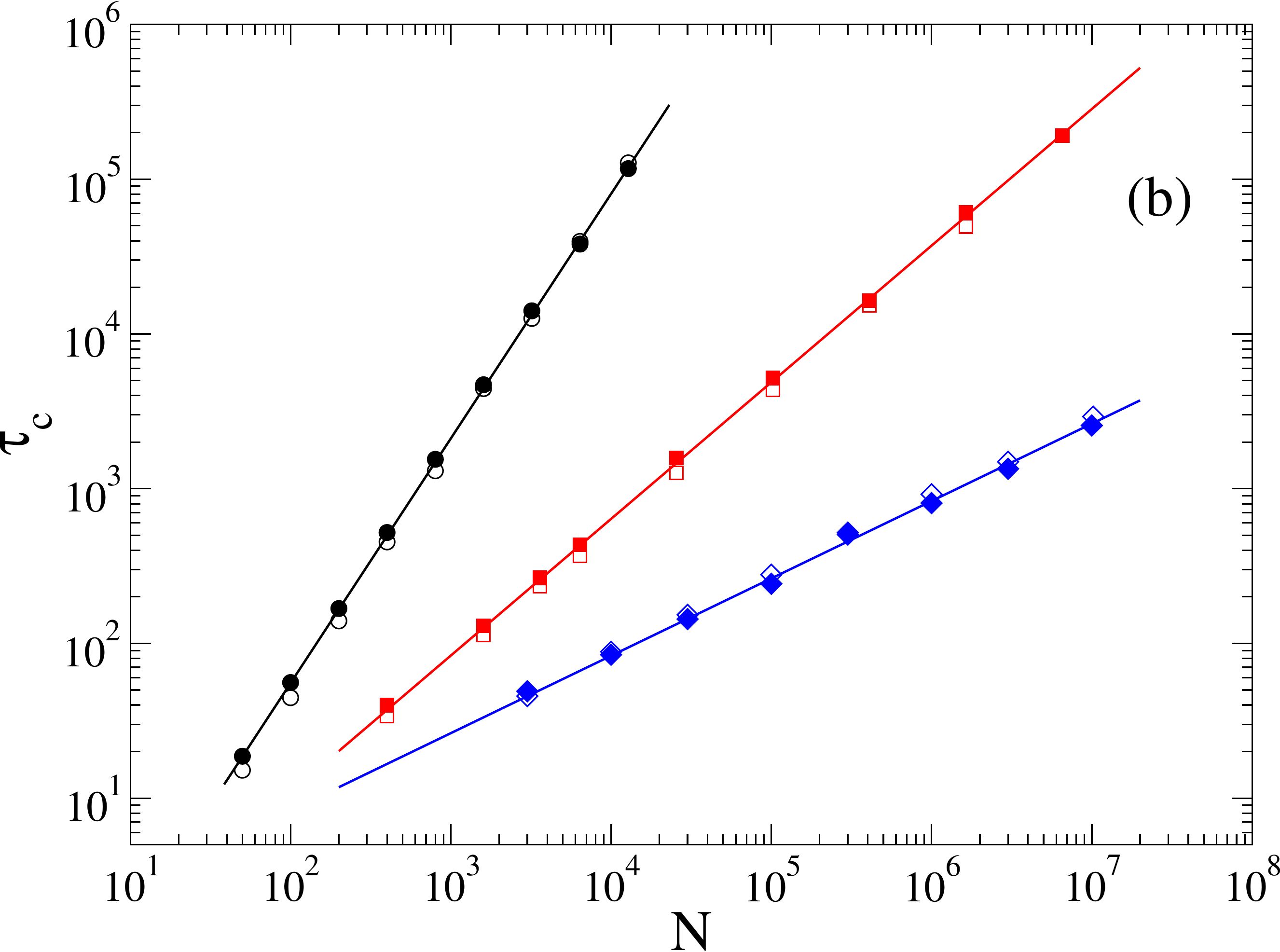}
\caption[ht]{(a) Lifespan of the critical SIS dynamics on lattices of dimensions
$d=1$, 2 and RRN ($d=\infty$). Solid lines represent the DP exponents predicted by the
SQS ($\tau_a\sim N^{z^*}$) and dashed ones represents RBC methods [$\tau_a\sim
N^{z^*(1-\delta)}$].  (b) Critical integrated correlation time
against size. Solid lines are PLs $\tau_c\sim N^{z^*}$ with the DP
exponents. }
\label{fig:tauaSISlattice}
\end{figure}

Figure~\ref{fig:tauaSISlattice}(a) compares the critical epidemic lifespan
$\tau_a$ as a function of the size $N$ for SQS and RBC methods. The DP scaling
laws $\tau \sim N^{z^*}$, with $z^*=1.5807$, $0.8830$, and 1/2 for $d=1$, 2, and
$\infty$, respectively, are confirmed for SQS method while the DP exponents
$z^*(1-\delta)=1.3285,~0.4852$, and 0 (logarithmic) are observed for RBC,
confirming the prediction of Eq.~\eqref{eq:tauaRBC}.
Figure~\ref{fig:tauaSISlattice}(b) compares the integrated autocorrelation times
for SQS and RBC at the critical point. One can see that the same scaling law
$\tau_c\sim N^{z^*}$ is found, showing that this quantity provides a correct
characteristic relaxation time for both methods.

\section{Numerical analysis on power-law degree distributed networks}
\label{sec:results}

We consider networks with PL degree distributions of the form $P(k) \sim
k^{-\gamma}$ generated by the uncorrelated configuration model
(UCM)~\cite{Catanzaro05}. The network is built associating the number of stubs
of each vertex according the distribution $P(k)$, randomly connecting stubs,
avoiding self- and multiple connections, and imposing the minimal degree $k_0=3$
and the structural upper cutoff $k_{c} = N^{1/2}$ in the degree distribution.
This procedure guarantees the absence of degree correlations for any value of
$\gamma>2$~\cite{Catanzaro05}. We analyze values of $\gamma$ for which SIS has
distinct behaviors on UCM networks of finite size~\cite{Ferreira16}. For
$\gamma<3$, SIS exhibits a single well resolved transition occurring at a
threshold that goes to zero in the thermodynamical limit while for $\gamma>3$
the model has localized active phases leading to multiple smeared transitions
for large networks~\cite{Mata15,Cota16}. In the case of CP, a sharp transition
occurring at a finite threshold is observed for any value of
$\gamma$~\cite{mata2014,Ferreira_quenched}. We consider only the case
$\gamma<3$ where the critical exponents depend on $\gamma$~\cite{RomuPRL2008}.

\subsection{$\gamma <3$}
\begin{figure}[ht]
\centering
 \includegraphics[width=0.8\linewidth]{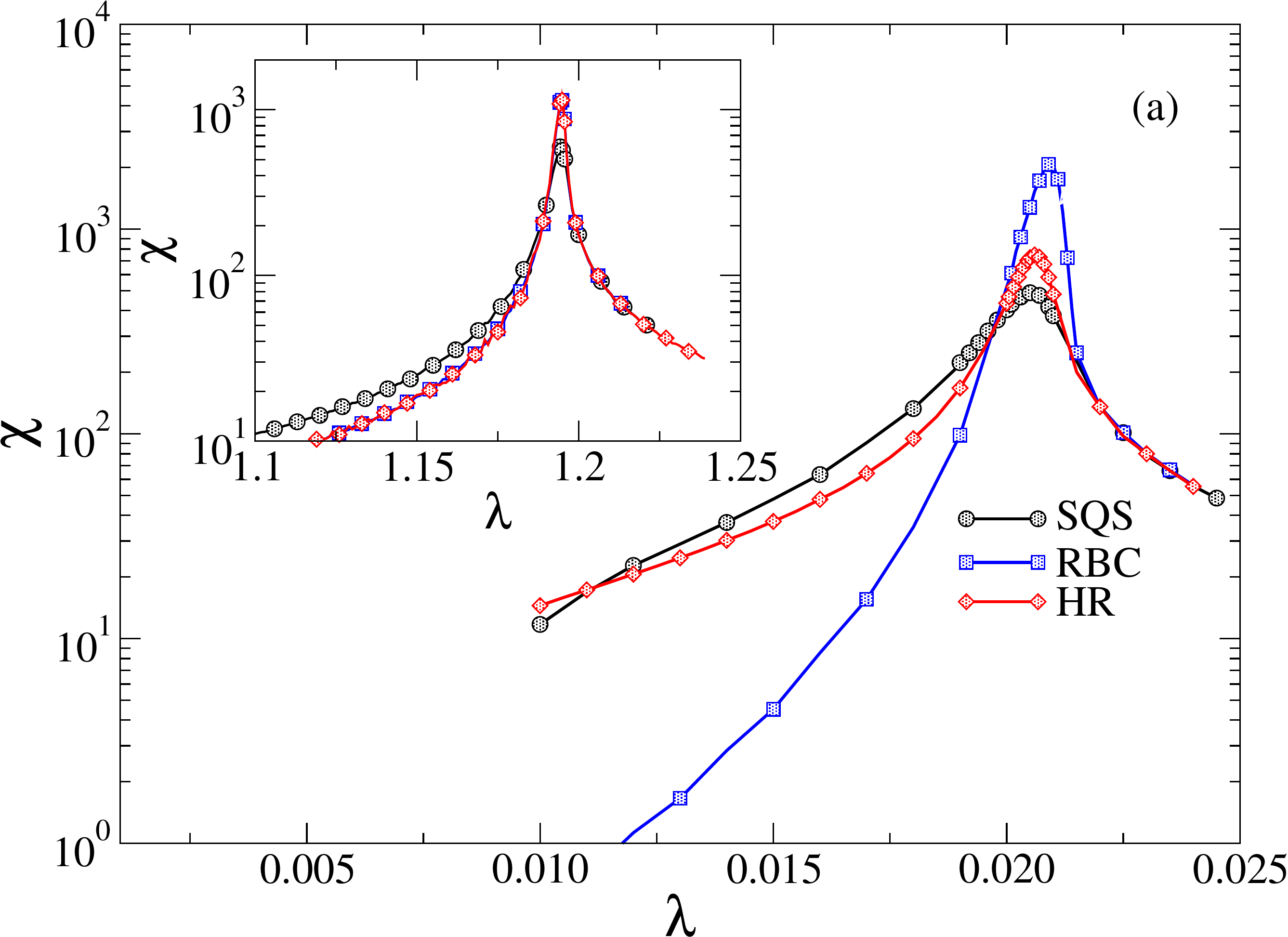}\\
 \includegraphics[width=0.8\linewidth]{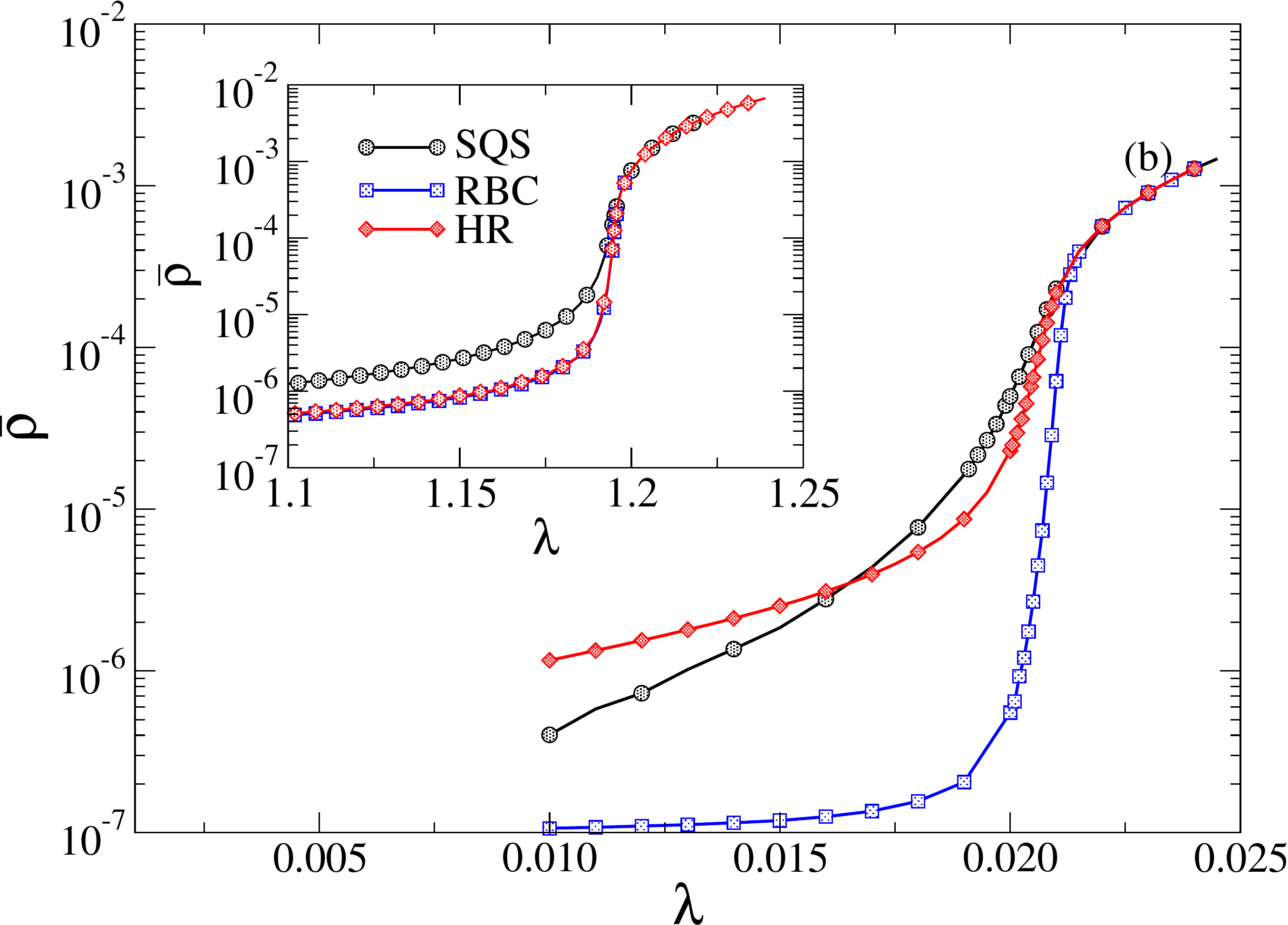}
 \caption{Comparison of QS quantities on SF networks. 
 (a) Susceptibility $\chi$ and (b) order parameter $\bar{\rho}$  in the
 QS regime as a  function of $\lambda$ for the SIS and CP (insets) models
 using different methods. The degree exponent is $\gamma=2.7$ and the
 network size is $N=10^7$.}
 \label{fig:rhochisis270}
\end{figure}
\begin{figure}[ht]
\centering
\includegraphics[width=0.85\linewidth]{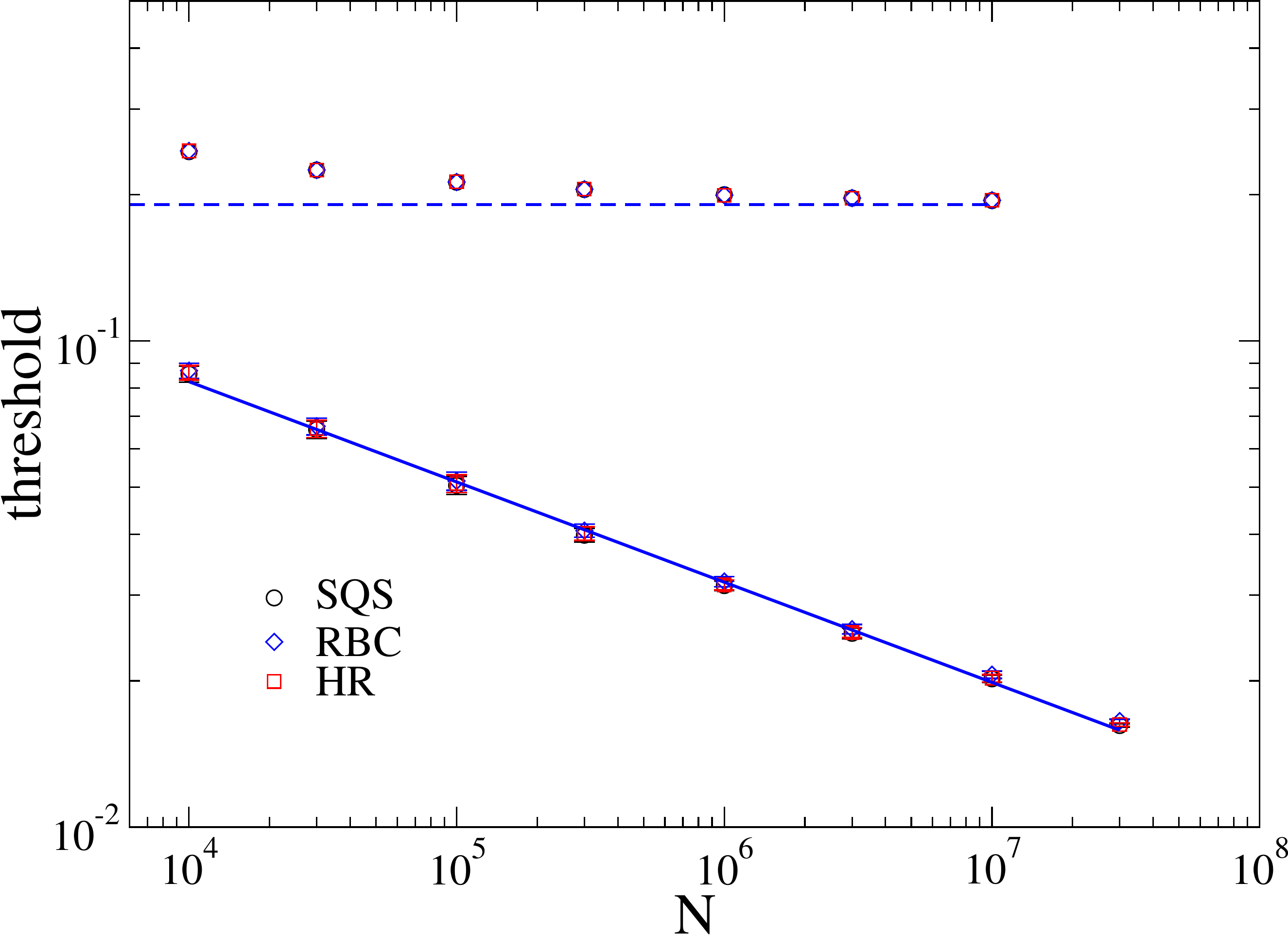}
 \caption{Effective thresholds $\lambda_{p}$  as a function of $N$ for the SIS
 (bottom curves) and CP (top curves) models obtained with  SQS, RBC, and HR
 methods. The CP threshold was subtracted by 1 to improve visibility. The solid
 line is a PL regression and the dashed one is the epidemic threshold of the
 CP in the limit of very large networks given in Ref.~\cite{mata2014}. The degree
 exponent is $\gamma=2.7$.}
 \label{fig:thresSISeCP270}
\end{figure}

A comparison of susceptibility curves obtained with RBC, SQS, and HR methods
for SIS and CP models is shown in Fig.~\ref{fig:rhochisis270}(a) for a network
with $N=10^7$ vertices. The three methods are equivalent for CP: The
transition is slightly less pronounced in SQS than RBC and HR, but the last two are
indistinguishable.  For SIS,  the susceptibility peak obtained with RBC is
evidently more pronounced than  with HR method, despite their algorithmic
similarity: The latter exhibits susceptibility close to the one of the SQS
method, indicating a smearing of the transition~\cite{Cota16} in SQS and HR when compared with
RBC. This phenomenon is also evidenced by the curves of density against
infection rate shown in Fig.~\ref{fig:rhochisis270}(b), in which  the subcritical
density in the SQS and HR methods are broader than in RBC.

The position of the susceptibility peaks is practically independent of method
used, as shown in Fig.~\ref{fig:thresSISeCP270}. In the case of SIS,  RBC
 provides peaks slightyly above other methods, a difference that is not perceivable
in this plot.  Fitting SIS data to a PL $\lambda_{p} \sim N^{-\phi}$, the
exponent is $\phi=0.20$ for the three methods. For CP the convergence to the
threshold value reported in Ref.~\cite{mata2014}, for this same network model, is
verified with the three methods.

\begin{figure}[ht]
	\centering
	\includegraphics[width=0.85\linewidth]{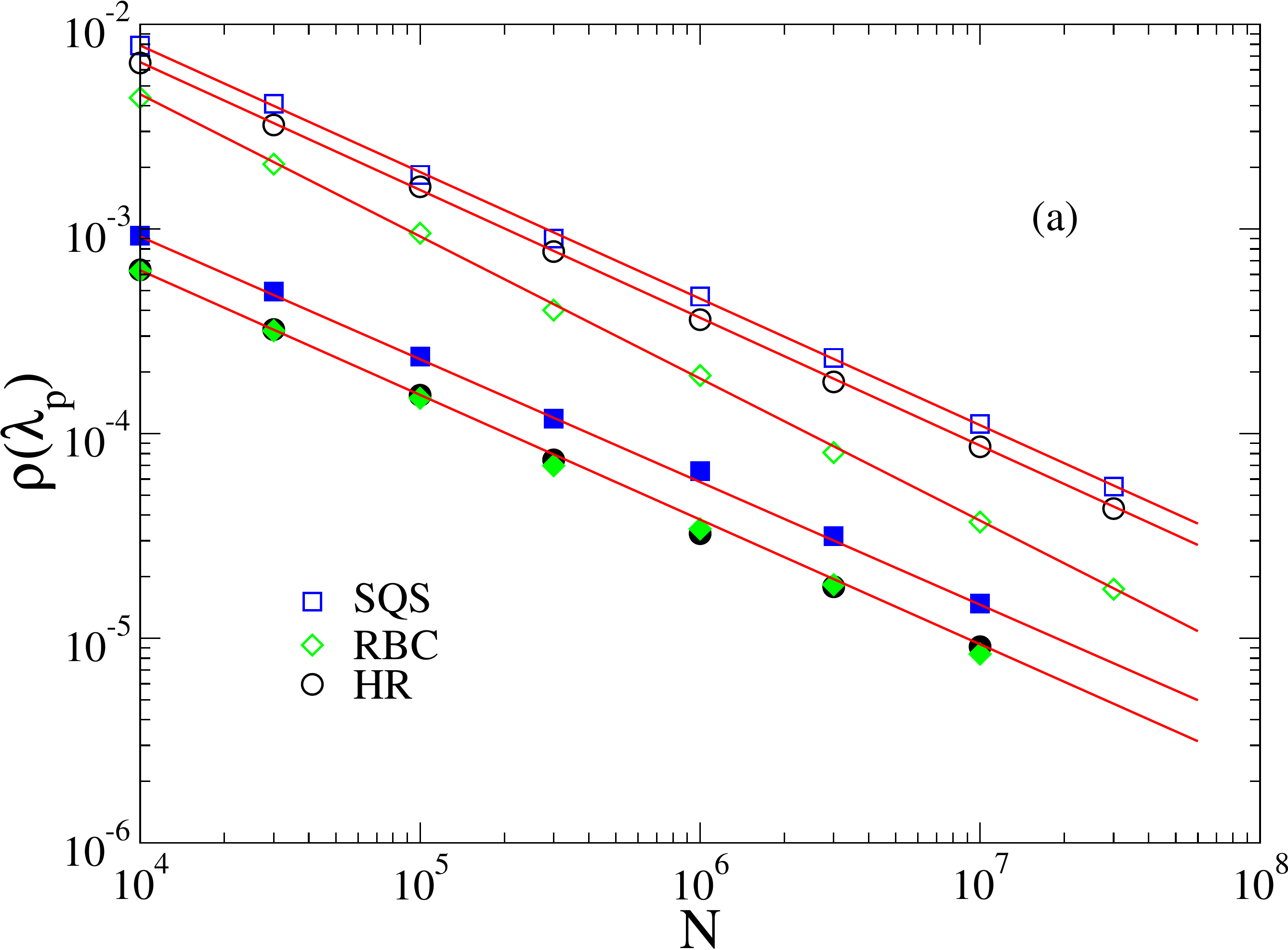}
	
	\includegraphics[width=0.85\linewidth]{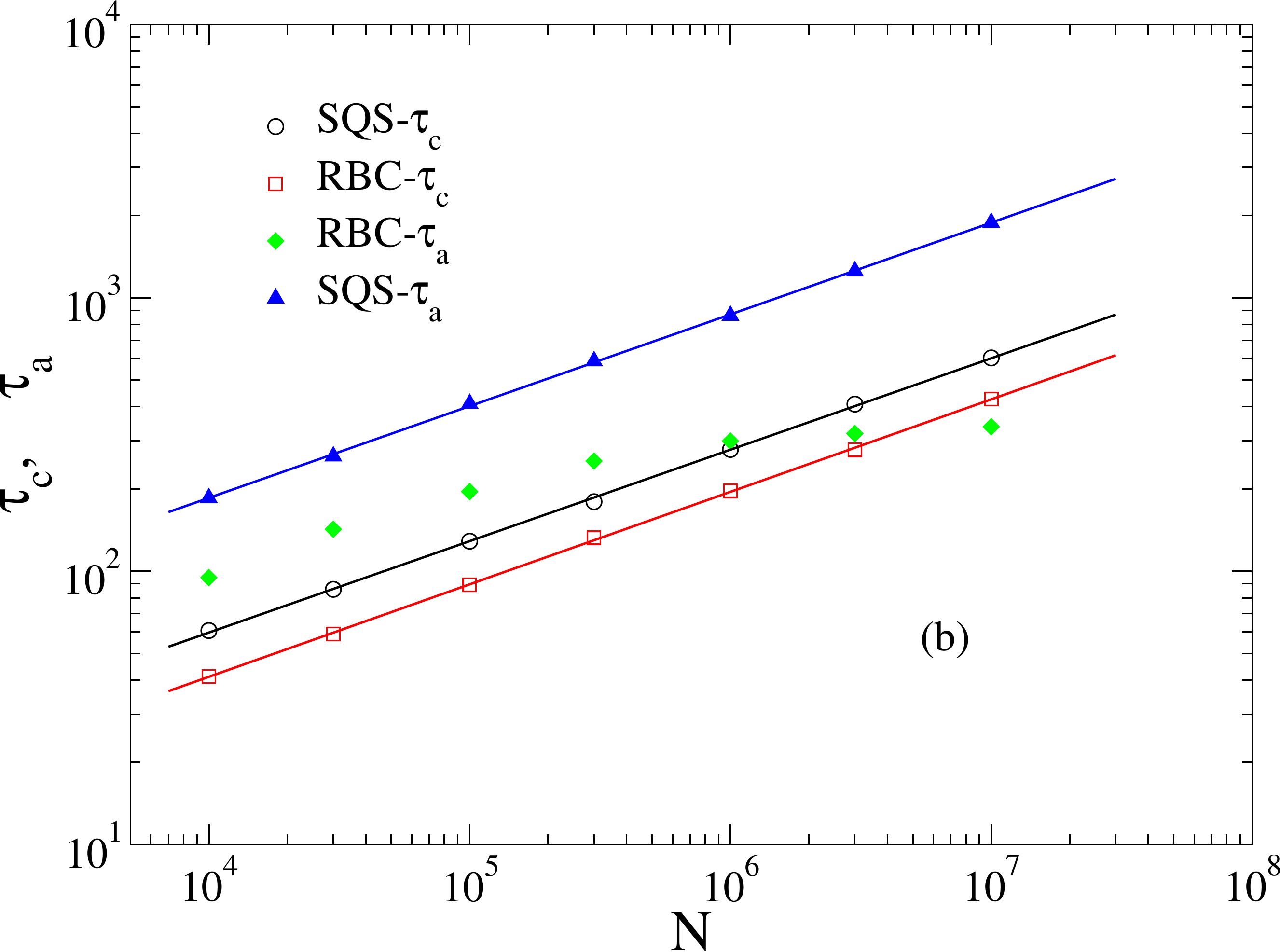}
	
	\caption{Finite size scaling of epidemic models on UCM networks  with
		$\gamma=2.7$. (a)  Critical density  for SIS (open symbols) and CP (filled
		symbols). The densities for CP were divided by a  factor 10 to improve the
		visibility. (b) FSS of the integrated correlation time ($\tau_c$) and time
		between attempts of visits to the absorbing state ($\tau_a$) for the SIS model. The
		SQS data for $\tau_a$ is divided by a factor $10^2$ to improve the visibility.  The
		lines represent PL regressions. The averages were performed over 10
		network realizations and the error bars are smaller than the symbols.}
	\label{fig:FSSg27}
\end{figure}

Figure \ref{fig:FSSg27}(a) shows the FSS of the order parameter $\rho$ evaluated
at the position of the susceptibility peak $\lambda_{p}(N)$~\cite{mata2014} for
$\gamma=2.7$. Assuming a scaling $\rho\sim N^{-\beta^*}$, the FSS of CP provides
$\beta^*=0.60(2)$ for RBC, HR, and SQS methods. Here the numbers in parentheses
represent uncertainties due to the regression. However, for the SIS model we
found $\beta^*=0.69(2)$ for RBC in contrast with and $\beta^*=0.61(2)$  obtained
for SQS and HR methods. Notice that the SQS exponents for SIS and CP are the
same within uncertainties. The FSS of the auto correlation time and epidemic lifespan for
SIS model are shown in Fig. \ref{fig:FSSg27}(b). As in the case of RRN networks, the
time $\tau_a$ increases logarithmically for RBC and HR (data not shown for the
latter) methods. On the other hand, the autocorrelation times provide the same
scaling for RBC, HR and SQS that agrees with the scaling of $\tau_a$ obtained
via the SQS method. Assuming a scaling $\tau\sim N^{z^*}$, we found the same
exponent $z^*=0.33(1)$ for $\tau_c$ in all methods and $\tau_a$ in SQS.
Equivalent results were obtained for CP with an exponent $z^*=0.40(1)$.

Let us shed some light on the difference between SIS simulations using RBC and
HR methods. In RBC, the probability of returning to the absorbing configuration
in the next step after the epidemic is restarted is $p_\mathrm{abs}=1
/(1+\lambda_c k)$, where $k$ is the degree of the vertex where the activity ended
and returned subsequently. Since the pre-absorbing state will, with large
probability, be in a vertex of low degree, which represents the great majority of
vertices in the network, and remembering that $\lambda_c(N)\rightarrow 0$ as
$N\rightarrow\infty$, the probability that the system falls back into the absorbing
state in the next step goes to 1. In the HR method, the particle returns to a
vertex of degree $k=k_c\sim \min[N^{1/2},N^{1/(\gamma-1)}]$ such that $\lambda_c(N) k\gg 1$ and the
probability to produce an outbreak with $n>1$ infected vertices goes to 1. So
this qualitatively explain why HR is similar to SQS rather than RBC since the
long-term regime of SQS is constituted by a large number of configurations with
$n\gg 1$, being many of them neighboring the hubs  and increasing the chance of
their reactivation. Note that in the CP dynamics, the system returns to the
absorbing state with probability $p_\mathrm{abs}=1/(1+\lambda_c)<1/2$
irrespective of the vertex degree and a new outbreak with $n>1$ has a finite
probability to happen in both RBC and HR. The autocorrelation is insensitive to
this detail and provides the same FSS scaling exponents for all methods.

\subsection{$\gamma>3$}

\begin{figure}[ht]
\centering
\includegraphics[width=0.95\linewidth]{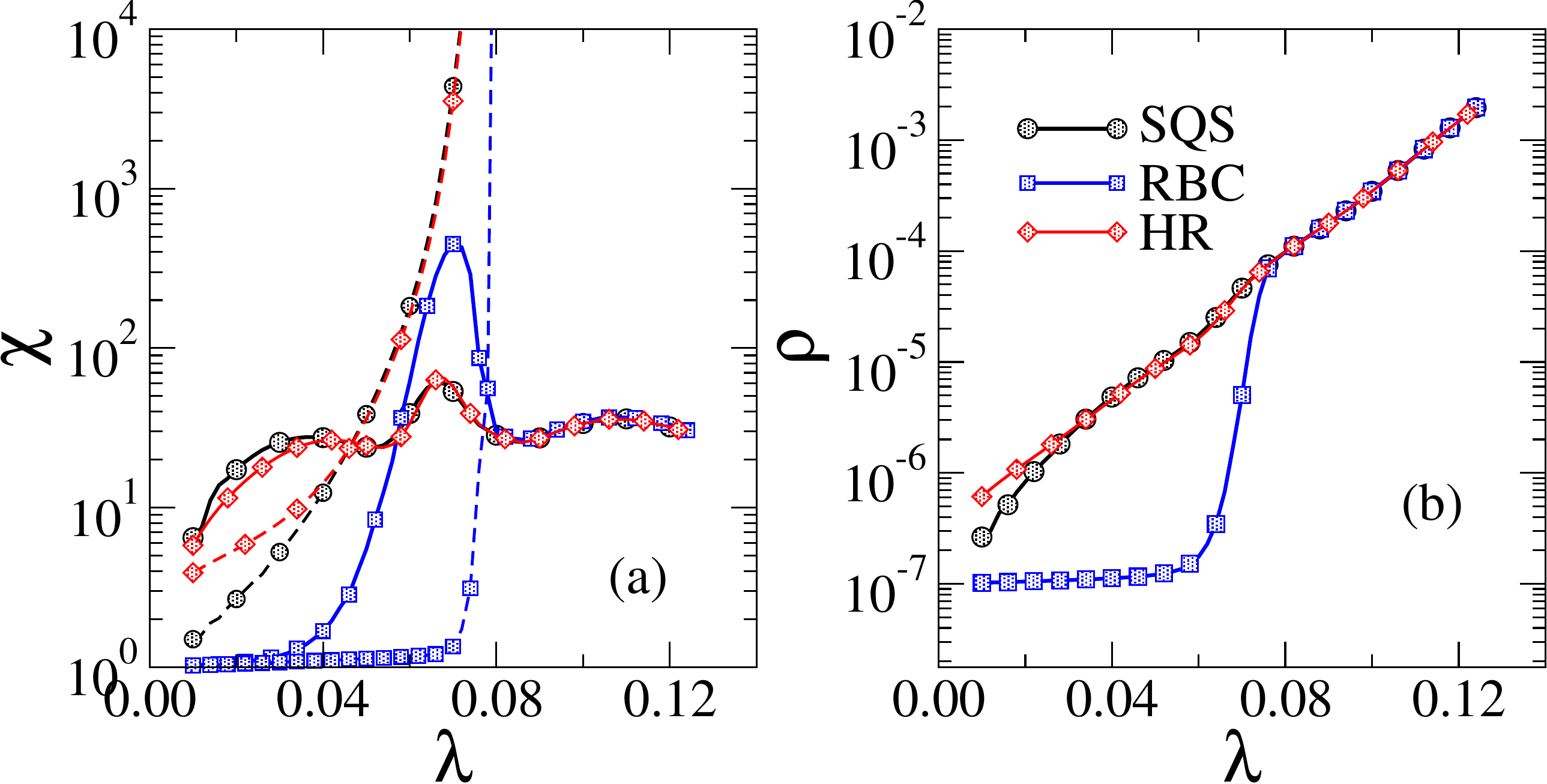}\\
\includegraphics[width=0.95\linewidth]{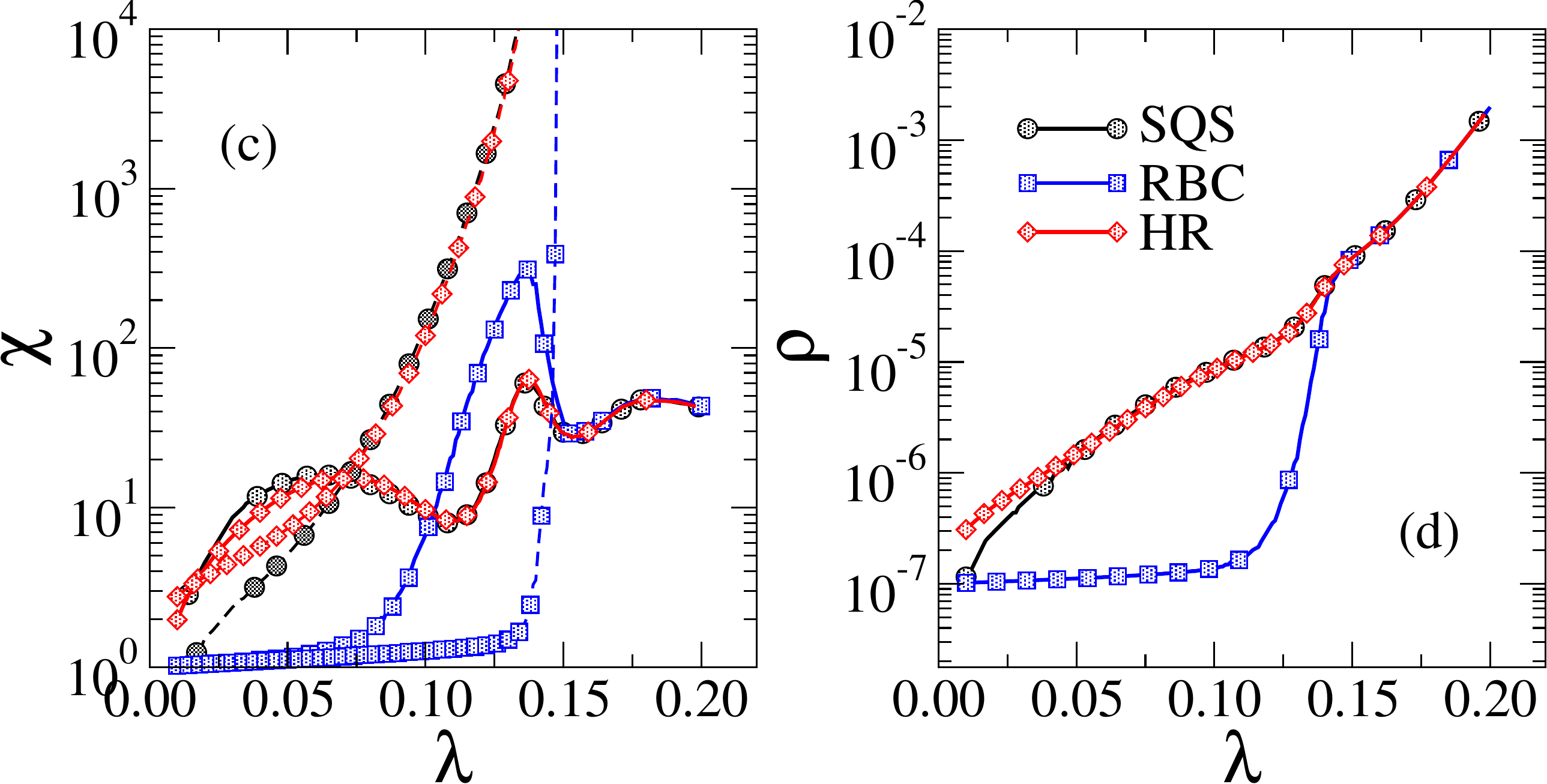}
\caption{Susceptibility $\chi(\lambda)$ and QS density $\rho(\lambda)$ curves
for the SIS model on UMC networks with (a), (b)$\gamma=3.5$, and (c),(d)
$\gamma=4.0$ using different QS methods. The dashed lines are the epidemic
lifespan $\tau_a=1/\bar{P}_1$. The network size is $N=10^7$.}
\label{fig:chirhovslbgs}
\end{figure}
For $\gamma>3.0$ the degree distribution of the UCM model has a finite variance
but presents some vertices with degree much larger than the rest of the
network, hereafter called outliers; see discussions in
Refs.~\cite{Mata15,Cota16}. These outliers generate localized metastable patches that can
be independently activated, manifested as multiple peaks in the susceptibility
of the SIS model on large networks simulated with the SQS
method~\cite{Ferreira12,mata2013pair,Mata15,deArruda}. Susceptibility curves
comparing the different methods for the same network realization are shown in
Fig.~\ref{fig:chirhovslbgs}.  A remarkable difference is that the peak observed
at small $\lambda$ in SQS curves, which is due to the activity localized in the
most connected vertex of the network~\cite{Ferreira12,Mata15}, is not observed
in the RBC but is in the HR method. The secondary peaks are associated to the
lifespan divergence~\cite{Mata15}. The SQS and HR methods are equivalent. The
activation of the most connected vertex is not captured by RBC method but it is
for the other two methods. A smearing of the
transition~\cite{Cota16,Vojta2006Rev,OdorsmearCP} in SQS and HR methods is more
evident than in the case $\gamma=2.7$. As expected, above the epidemic
threshold, the methods become equivalent since the absorbing state was never
visited in the simulations. It is worth mentioning that the multiple transitions
are not artifacts of the SQS simulations, and are also observed with the HR and
RBC methods.

\begin{figure}[ht]
\centering
 \includegraphics[width=0.95\linewidth]{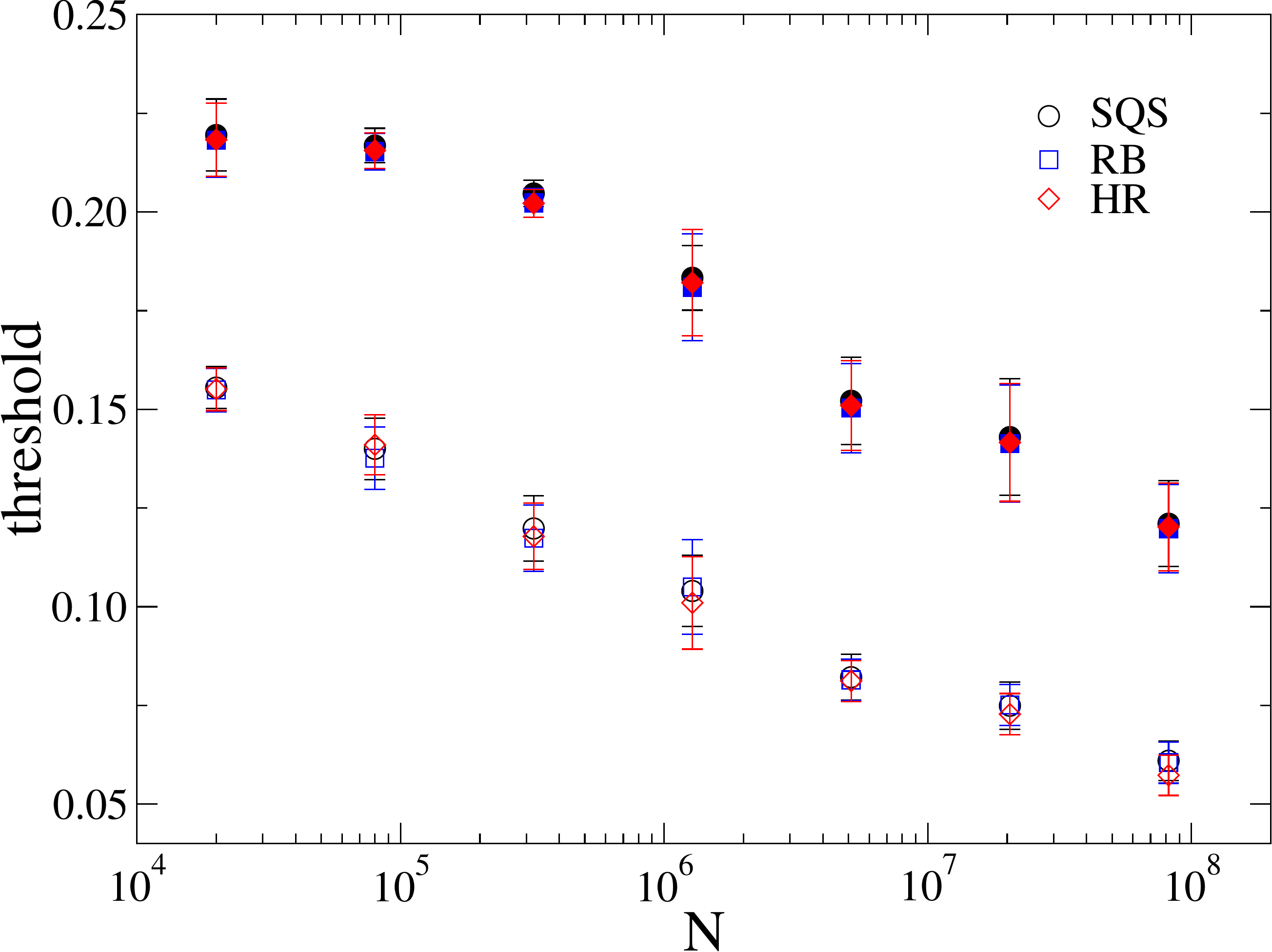}\\
 \caption{Effective thresholds  as a function of the system size for the SIS model
 	on UCM networks with  $\gamma=3.5$ (open symbols) and $\gamma=4.0$ (filled
 	symbols) using different QS methods.}
 \label{fig.tresmetodos_sis350e400-thresholds}
\end{figure}

Figure~\ref{fig.tresmetodos_sis350e400-thresholds}  shows the scaling of the
epidemic threshold of SIS model for $\gamma=3.5$ and $\gamma=4.0$ as a function
of the network size. One can see that the size dependences are remarkably
similar for the three methods and the differences are inside uncertainties.

\subsection{Comparison between WEF and RBC methods}

\begin{figure}[th]
\centering
\includegraphics[width=0.95\linewidth]{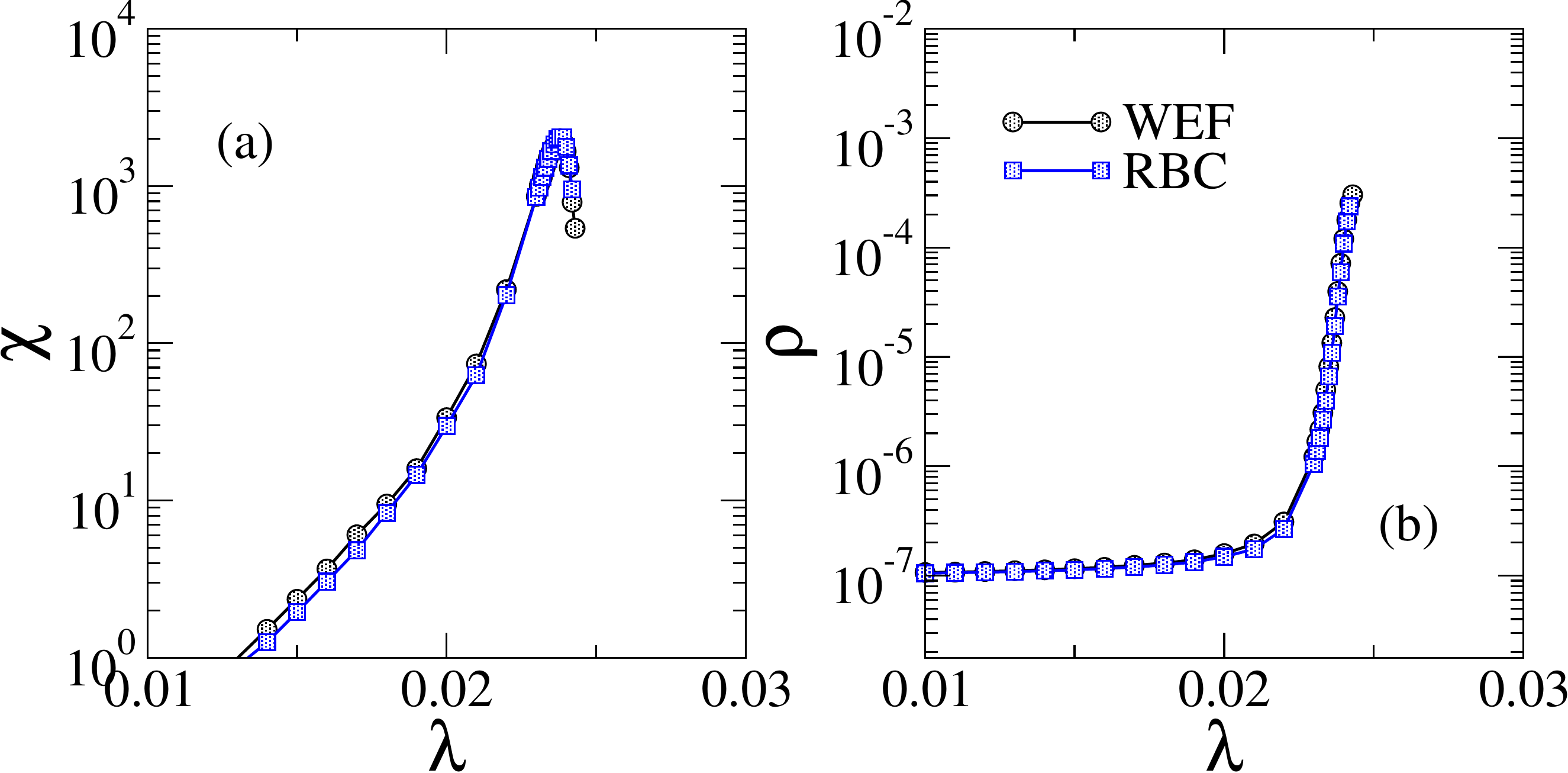}\\
\includegraphics[width=0.95\linewidth]{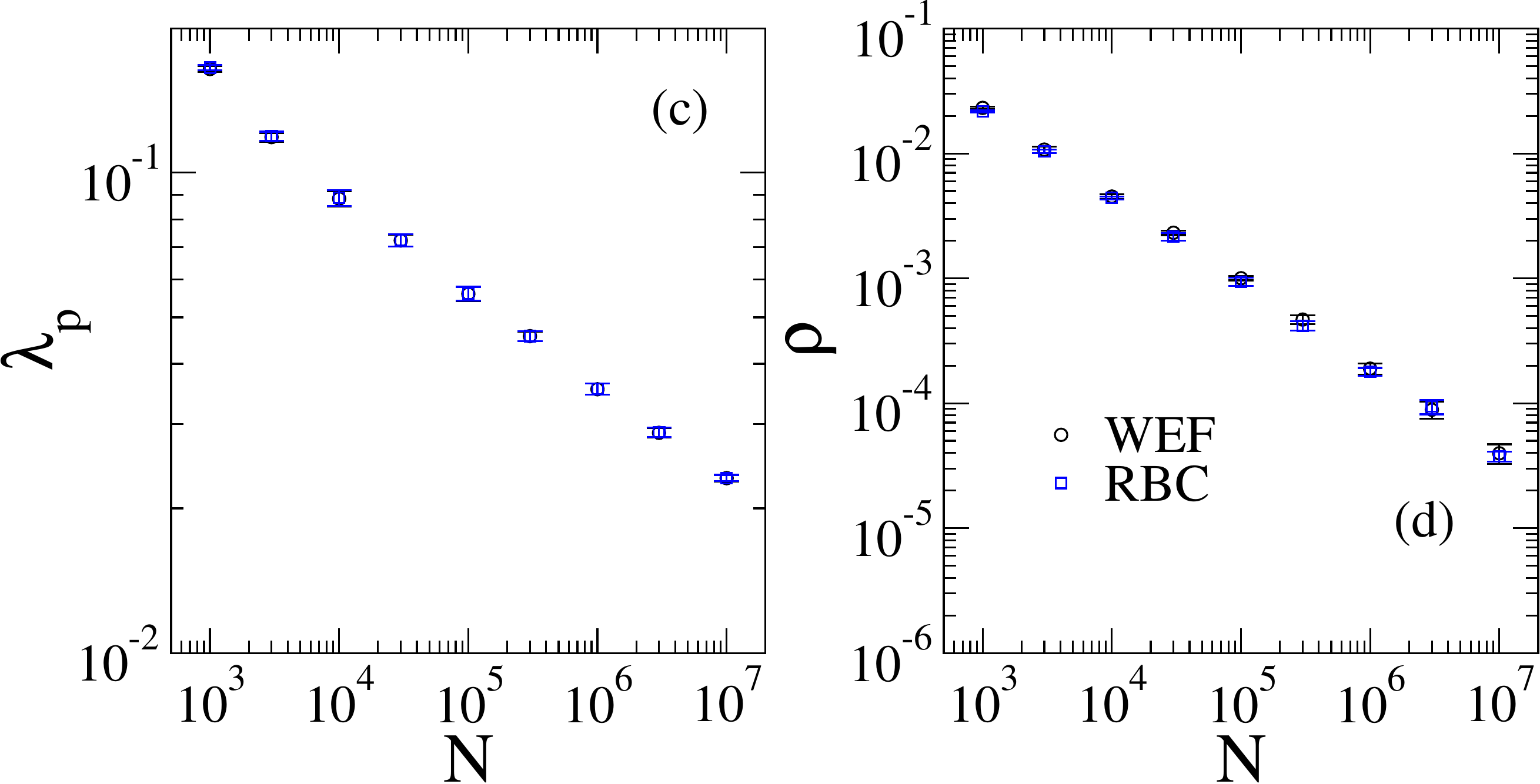}\\
\caption{Equivalence of RBC and WEF methods for SIS model on UCM networks with
degree exponent $\gamma=2.75$. Top panels show the (a) susceptibility and (b)
density of infected vertices against infection rate. Bottom panels show the FSS
of (c) epidemic threshold and (d) critical density.}
\label{fig:rbcvsefg27}
\end{figure}

In the limit of the weak external field (WEF), the creation events rarely occur
in the active phase and therefore it is equivalent to RBC method in regular
graphs~\cite{gunnar}. However, some differences are eligible in highly
heterogeneous networks since pre-absorbing states are usually configurations
with infection near or in a hub. This could enhance the chance of epidemic outbreak
occurrence in RBC method because the epidemic would restart nearer a hub than
in WEF method, where it returns to a randomly selected vertex. We performed
simulations using a external field $f=N^{-1.25}$. A comparison of the methods for SIS in
Fig.~\ref{fig:rbcvsefg27} shows the equivalence between them.

\section{Conclusions}
\label{sec:conclu}

The interplay between structural properties of complex networks and evolution of
epidemic processes on the top of them constitutes a fundamental problem in
network science~\cite{RMPreview2014}. Distinct theories have been developed to
address central questions such as the position or existence of epidemic
thresholds~\cite{RMPreview2014,pv01a,Ganesh05,Mieghem11,RomuPRL2010,Ferreira16,boguna2013,Gleeson}, localization and delocalization of epidemic phases~\cite{Goltsev12,Odor13,Lee2013,MunozPRL2010}, and the critical exponents ruling the epidemics around the threshold~\cite{RomuPRL2006,RomuPRL2008,HongPRL2007,mata2014}. In this framework, numerical simulations are fundamental tools in the validation of theories and conjectures~\cite{pv01a,RomuPRL2010,Ferreira12,Ferreira_quenched,mata2013pair,Gleeson} as well as in the setting up of new physical and analytical insights~\cite{boguna2013,Ferreira16,OdorsmearCP}.

Simulations near the epidemic threshold constitute a challenge since in finite
networks an absorbing state, in which the epidemic is eradicated, will always
be reached due to the finite number of accessible
configurations~\cite{marro1999npt}. Aiming at investigating epidemic processes
with steady states via stochastic simulations, one needs to resort to the QS
approaches, which suitably handle the absorbing states, along with a finite-size analysis. In these approaches a perturbation of the original dynamics, which is neglibigle in the thermodynamical limit of the active phase, is introduced. We
investigated distinct QS methods: The SQS has sampling constrained to active
configurations; the RBC where the dynamics returns to the configuration that it
was immediately prior the visit to the absorbing state; the HR method, where the
epidemics is restarted in one of the most connected vertices of the network; and
a weak external field (WEF) that introduces spontaneous infection. Two distinct epidemic
models with active steady state were considered, the CP~\cite{marro1999npt} and the
SIS~\cite{RMPreview2014} models, whose phase
transitions have different natures~\cite{Ferreira16}. 

We observed that all methods are equivalent for CP, providing the same epidemic
thresholds and FSS exponents of the critical QS quantities. For SIS, the same
thresholds are obtained for all methods but the FSS of the critical density
provides scaling exponents for RBC and WEF different from SQS and HR methods.
Also, RBC and WEF do not capture epidemic activity localized in the most
connected vertex of the network. So, if one wishes an analysis rid of localized
epidemics, RBC method is indicated but if one also needs to resolve localization
the SQS or HR are more appropriated. The SQS is theoretically well
grounded~\cite{blanchet,dickman2002quasi,Ferreira_annealed} but it is
algorithmically more complicated and computationally less efficient than the
other investigated methods. 

An advantage of the SQS method is that it  provides an epidemic
lifespan  proportional to the
characteristic relaxation time in both subcritical and critical
regimes~\cite{Mancebo2005}, but it is infinite in the active phase. Moreover,
for the other investigated methods the epidemic lifespan 
 does not correspond to the characteristic
relaxation time. In order to overcome these difficulties we analyzed the
auto correlation time of the QS series for the different methods.
Auto correlation provides the characteristic relaxation time, including in the
supercritical regime. We found the same FSS exponents for all investigated
methods irrespective of the network and epidemic models considered. In
particular, applying the auto correlation method to regular lattices of
dimension $d=1, 2,$ and 3 as well as RRN ($d=\infty$), the directed percolation
exponents were obtained.

\begin{acknowledgments}
  This work was partially supported by the Brazilian agencies CNPq, FAPEMIG, and
  CAPES.  S.C.F. thanks with Romualdo Pastor-Satorras for discussions during visits to
  UFV supported by the program \textit{Ci\^encia sem Fronteiras} - CAPES (Grant
  No.  88881.030375/2013-01).
\end{acknowledgments}

%

\end{document}